\documentstyle[aps,preprint,epsf]{revtex}

\begin{document}

\draft

\title{Jamming and Fluctuations in Granular Drag}
\author{I. Albert$^{1}$, 
P. Tegzes$^{1,2}$, B. Kahng$^{1,3}$, R. Albert$^1$, J. G. Sample$^1$, 
M. Pfeifer$^{1}$, \\ A.-L. Barab\'asi$^{1}$, T. Vicsek$^{2}$, 
and P. Schiffer$^{1, *}$ }

\address{
$^1$ Department of Physics, University of Notre Dame, Notre Dame, IN 46556\\
$^2$ Department of Biological Physics, E\"otv\"os University, Budapest 1117, Hungary\\
$^3$ Department of Physics, Konkuk University, Seoul 143-701, Korea
}

\date{\today}
\maketitle

\begin{abstract}
We investigate the dynamic evolution of jamming in granular media through
fluctuations in the granular drag force.  The successive collapse and
formation of jammed states give a stick-slip nature to the fluctuations
which is independent of the contact surface between the grains and the
dragged object -- thus implying that the stress-induced collapse is
nucleated in the bulk of the granular sample. We also find that while 
the fluctuations are periodic at small depths, they become
"stepped" at large depths, a transition which we interpret as 
a consequence of the long-range nature of the force chains.
\end{abstract}


\tightenlines
Materials in granular form are composed of many solid particles that
interact only through contact forces. In a granular pile, 
the strain resulting from the grains' weight combines with
the randomness in their packing to constrain the motion of individual
grains. This leads to a "jammed" state \cite{MECates} which also characterizes 
a variety of other frustrated physical systems, such as dense 
colloidal suspensions and spin glasses \cite{MECates,AJLiu}. 
Not surprisingly, the dynamics of granular materials, while in many ways 
analogous to those of fluids, are in fact quite different due to this 
frustration of local motion \cite{HMJaeger}. The effects  of this jamming 
are also manifested in static properties, leading to inhomogeneous
stress propagation through force chains of strained grains 
\cite{CLiu,BMiller,XJia,EKolb,ALDemirel,SNCopper,AVTkachenko,LVanel} 
and arch formation \cite{PClaudin}. 
Although jamming in granular materials has previously been discussed
in the context of the gravitational stress induced by the weight of the
grains, it can result from any compressive stress. 
For example, a solid object being pulled slowly through a granular medium  
is resisted by local jamming, and can only advance with large scale 
reorganizations of the grains.
The granular drag force originates 
in the force needed to induce such reorganizations, 
and thus exhibits  strong fluctuations which qualitatively distinguish 
it from the analogous drag in fluids \cite{RAlbert}.

In this Letter we investigate the dynamic evolution of jamming in granular
media through these fluctuations in the granular drag force.  The
successive collapse and formation of  jammed states give a stick-slip
character to the force with a power spectrum proportional to $1/f^2$.  
We find that the stick-slip process does not
depend on the contact surface between the grains and the dragged object, and
thus the slip events must be nucleated in the bulk of the grains opposing 
its motion. While the fluctuations are remarkably periodic for small depths, they
undergo a transition to "stepped" motion at large depths. 
These results point to the importance of the long-range nature of the 
force chains to both the dynamics of granular media and the strength of 
the granular jammed state.
                                                           
The experimental apparatus, shown in Fig.\ \ref{Fig1}a, consists of a 
vertical steel cylinder of diameter $d_c$ inserted to a depth $H$ in a 
bed of glass spheres \cite {Jaygo} moving with constant speed \cite{RAlbert}. 
The cylinder is attached to a fixed force cell \cite{Apparatus}, 
which measures the force $F(t)$ acting on the cylinder
as function of time. 
The bearings on the cylinder's support structure 
allow the cylinder to advance freely only in the direction of
motion so that the force cell alone is opposing the drag force from the
grains. We incorporate a spring of 
known spring constant, $k$, between the cylinder and the force cell -- 
choosing $k$ (between 5 to 100 N/cm) so that this spring dominates 
the elastic response of the cylinder and all other
parts of the apparatus. 
We vary the 
speed ($v$) from 0.04 to 1.4 mm/s, the depth of insertion ($H$) from 20 to 
$190$ mm, and the cylinder diameter ($d_c$) from 8 to 24 mm, studying grains
of diameter ($d_g$) 0.3, 0.5, 0.7, 0.9, and 1.1 mm \cite{Plots}. 
The force is recorded at 150 Hz and the response time of the force cell and 
the amplifier are $< \, 0.2$ ms. 

Consistent with earlier results \cite{RAlbert}, 
we find that the average drag force
on the cylinder is independent of $v$ and $k$, and is given by 
$\overline{F} = \eta\rho gd_cH^2$,
where $\eta$ characterizes the grain properties (surface friction, packing
fraction, etc.), $\rho$ is the density of the glass beads, and $g$ is
gravitational acceleration. 
As shown in Fig.\ \ref{Fig2}, however, $F(t)$ is not constant, 
but has large stick-slip fluctuations consisting
of linear rises associated with a compression of the
spring and sharp drops associated with the collapse of the 
jammed grains opposing the motion.
The linear rises in $F(t)$ correspond to the development 
of an increasingly compressed jammed state of the grains opposing the motion.
We find that, independent of the
depth, the slopes of the rises are given by 
$\frac{1}{v}dF/dt = k$ for all springs with $k<100$ N/cm, confirming
that the spring dominates the elasticity of the apparatus.
This result also implies that during the rises the jammed grains opposing 
the cylinder's motion do not move relative to each other or the cylinder.
The power spectra, $P(f)$ (the squared amplitudes of the Fourier 
components of $F(t)$), are independent of 
both the elasticity of the apparatus and the rate of motion, 
so that the scaled spectra $kvP(f) \mbox{ vs. } f/kv$ collapse 
in the low frequency regime ($f < 10$ Hz)\cite{BMiller}. 
This indicates that the fluctuations reflect intrinsic properties 
of the development and collapse of the jammed state rather than 
details of the measurement process.
The power spectra  also exhibit a distinct power law, $P(f) \propto f^{-2}$, 
over as much as two decades in frequency (Fig.\ \ref{Fig3}) a phenomenon 
which has been reported in other stick-slip processes 
and is intrinsic to random sawtooth signals \cite{Couette}.

During each fluctuation the force first rises to a local 
maximum value ($F_{max}$), and then drops sharply (by an amount $\Delta F$), 
corresponding to a collapse of the jammed state. The force from the cylinder propagates 
through the medium via chains of highly strained grains, and a 
collapse occurs when 
the local interparticle forces somewhere along one of the chains exceed a 
local threshold. The corresponding grains then slip relative to each other,
which in turn nucleates an avalanche of grain reorganization to relieve 
the strain. This allows the cylinder to advance relative 
to the granular reference frame, with a corresponding decompression 
of the spring and a drop in the measured force $F(t)$ \cite{Precursors}.
The interparticle forces within the force chains are
largest at the cylinder's surface, where the chains originate, and
their magnitude decreases as we move away from the cylinder and the force chains
bifurcate. Consequently, one might expect that the reorganization is nucleated among 
grains in contact with the cylinder, but
we find no change either in $\overline{F}$
or in the fluctuations when we vary the coefficient of friction between the grains
and the cylinder by a factor of 2.5 (substituting a teflon-coated
cylinder for the usual steel cylinder). As demonstrated in the inset to
Fig.\ \ref{Fig3}, the power spectra are also unchanged even by substituting a
half-cylinder (i.e. a cylinder bisected along a plane through its axis and 
oriented so that plane is normal to the grain flow) for a full cylinder 
of the same size, indicating that the geometric 
factors do not play a significant role either.  
These results indicate that the fluctuations are not determined by 
the interface between the dragged object and the medium, but rather 
that the failure of the jammed state is nucleated  {\em within the 
bulk of the medium}. In this respect, the fluctuations are rather 
different from either ordinary frictional stick-slip processes 
which originate at a
planar interface between moving objects, or the motion of a frictional 
plate on top of a granular medium \cite{SNasuno}.

A striking feature of the data is that 
the fluctuations change character 
with depth. For $H < H_c \approx 80$ mm the fluctuations 
are quite periodic, i.e. 
$F(t)$ increases continuously to a nearly constant value of $F_{max}$ 
and then collapses with a nearly constant drop of 
$\Delta F$ (Fig.\ \ref{Fig2}).
As the depth increases, however, we observe a change in $F(t)$ 
to a "stepped" signal: 
instead of a long linear increase followed by a 
roughly equal sudden drop, $F(t)$ rises in small linear increments to increasing
values of $F_{max}$, followed 
by small drops (in which $\Delta F$ is on average smaller than the rises), 
until $F_{max}$ reaches a characteristic high value, at which point 
a large drop is observed. 
This transition from a periodic to a "stepped" regime
is best quantified  in Fig.\ \ref{Fig4}, 
where we plot the depth dependence of $\overline{\Delta F}$ 
and the relative standard deviation of $\Delta F$, $\sigma_n =
\sigma_{\Delta F}/\overline{\Delta F}$. 
In the periodic regime,  $\overline{\Delta F}$ rises due to the increase in 
$\overline F$.
As the large uniform rises of the periodic regime are broken up by the small
intermediate drops, however, $\overline{\Delta F}$ shows a local minimum and
$\sigma_{\Delta F}/\overline{\Delta F}$ increases drastically,  
saturating for large depths. 
The transition 
is also observed in the power spectra 
as shown in the upper inset to Fig.\ \ref{Fig4}. 
For low depths the power spectra display a distinct peak
characteristic of periodic fluctuations, but 
these peaks are 
suppressed for $H>H_c$ in correlation with the changes in  
$\sigma_{\Delta F}/\overline{\Delta F}$ and the 
qualitative character of $F(t)$. 

The transition from a periodic to a "stepped" signal is rather
unexpected, since it implies qualitative changes in the 
failure and reorganization process as $H$ increases and
the existence of a critical depth, $H_c$. 
An explanation for $H_c$ could be provided by 
Janssen's law \cite{Janssen} which states that the average pressure (which 
correlates directly with the local failure process) should become 
depth independent below some critical depth in containers with
finite width. 
This should not occur in our container, however, which has a 
diameter of 25 cm, much larger than $H_c$.
Furthermore we see no deviation in the behavior of $\overline F(H)$ 
from $\overline F\propto H^2$, which depends on the pressure increasing linearly with
the depth (Fig.\ \ref{Fig4} lower inset).

In order to account for the observed transition, we must inspect how the
force chains originating at the surface of the cylinder nucleate the
reorganizations.  The motion of the cylinder attempting to advance
is opposed by force chains that start at the cylinder's surface and
propagate on average in the direction of the cylinder's motion.
These force chains will terminate rather differently depending on the
depth at which they originate, as shown schematically in Fig.\
\ref{Fig1}b.  For small $H$, some force chains will terminate on the top
surface of the granular sample and the stress can be relieved by a slight
rise of the surface. Force chains originating at large depths, however,
will all terminate at the container's walls. Since the wall does not
permit stress relaxation, the grains in these force chains will be more
rigidly held in place. According to this picture, $H_c$ corresponds to the   
smallest depth for which all force chains terminate on the wall.
When the cylinder applies stress on the medium,
the force chains originating at small H ($H<H_c$) reduce their strain
through a microscopic upward relaxation of chains ending
on the free surface. By contrast, the higher rigidity of force chains
originating at $H>H_c$ impedes such microscopic relaxations.  Thus a
higher
proportion of the total force applied by the cylinder will be supported by
those force chains, enhancing the probability of a local slip event
occuring at high depths. Such a slip event  would not necessarily
reorganize the grains at all depths (for example the grains closer to the surface
may not be near the threshold of reorganization), thus the 
slip event might induce only a local reorganization and a small drop in $F(t)$.
The large drops in $F(t)$ would occur when force chains at all
depths are strained to the point where the local forces are close to the threshold  
for a slip event. This scenario also explains why $\overline{F} (H)$ does
not change at $H_c$, since $\overline F$ is determined by the collective
collapse of the jammed structure of the system.  

According to this picture, the transition
is expected at smaller depths in smaller containers since 
the force chains would terminate on the walls 
sooner (seen Fig.\ \ref{Fig1}b). Indeed,
as we show in Fig.\ \ref{Fig4}, the transition does occur at a
depth approximately 20 mm smaller when the measurements are performed
in a container 2.5 times smaller (with diameter of 100 mm). 
Furthermore, we fail to observe the periodic
fluctuations in any grains with diameters 1.4 mm or larger \cite{Beads}, 
which is consistent with the suggested mechanism, since larger grains 
correspond to a smaller effective system size. 

It is interesting to compare our results with those of Miller et al. \cite{BMiller} 
who studied fluctuations in the normal stress at the bottom of a sheared
granular annulus.  Those fluctuations were also independent of the rate of
the motion, and demonstrated the long-range nature of the vertical force
chains.  In the present experiments, we confirm that force
chains originating from a horizontal stress are also long range through our 
observation of the transition at depth $H_c$.  Since for small 
grains ($d_g\leq 1.1$ mm) $H_c$ has no measurable 
dependence on grain size, we find in agreement with Miller et al. that the nature of the
force chains is not strongly dependent on grain size.  Our observations
also shed light on the implications of the long-range force chains for granular
dynamics and the nature of jamming in granular materials.  The crossover at
$H_c$ suggests that drag fluctuations in an infinitely wide container would
be periodic, but that the finite size of a real container destroys the
periodicity.  In other words,  the  finite size of a container relative to
these chains reduces the strength of jammed granular states within the
container.   These results point to the need for a better 
understanding of the detailed dynamics of  force chains  -- both how they
form when stress is applied to a granular medium and how they disperse
geometrically from a point source of stress --in order to gain an 
understanding  of slow granular flows.

We gratefully acknowledge the support of the Petroleum Research
Foundation administered by the ACS, 
the Alfred P. Sloan Foundation, and NSF grants PHYS95-31383 and DMR97-01998.

\begin{figure}
\caption{
a.) An enlarged view of the apparatus used for measuring the drag force. The
grains were contained within a 25 cm diameter rotating bucket as described in
detail previously \protect\cite{RAlbert}.
b.) Schematic illustration of the force chains originating at the surface of an 
object dragged to the right in a finite container, where  
$H_c$ corresponds to the depth at which the force chains all terminate at 
the walls of the container. Of course, this picture is highly simplified 
since the actual force chains bifurcate and follow non-linear paths.
}
\label{Fig1}
\end{figure}

\begin{figure}
\caption{
The characteristic fluctuations in the drag force at 4 different values
of $H$ for $d_c=10$ mm.  Note the transition from purely periodic 
fluctuations $H\leq 60$ to stepped fluctuations with increasing depth $H\geq 100$. 
}
\label{Fig2}
\end{figure}

\begin{figure}
\caption{
The power spectrum of the fluctuations in the drag force taken for 
$d_g = 1.1$ mm and $d_c = 16$ mm at depth $H=60$ mm with 
$v= 0.05$ mm/s and $k = 5$ N/cm  to increase the 
dynamic range. Note that in both the periodic and the stepped regimes 
the spectrum has a long $f^{-2}$ regime as shown in the upper inset
(shifted vertically for clarity) ($d_c=16$ mm).
The lower inset ($d_g=1.1$ mm, $d_c=19$ mm) shows that the 
power spectrum does not change when a half-cylinder is substituted for a full cylinder.
}
\label{Fig3}
\end{figure}

\begin{figure}
\caption{
The transition from periodic to stepped fluctuations as shown through 
the  magnitude of the average drop $\overline{\Delta F}$, for two 
different container sizes: circles - large container ($D =$25mm),
triangles - small container ($D=10$ mm).
The upper inset shows the relative magnitude of the standard 
deviation  $\sigma_n = \sigma_{\Delta F}/\overline{\Delta F}$. 
The transition occurs at smaller $H_c$ in the smaller container.
($d_c= 10$ mm) The lower inset shows the depth dependence 
of the average drag force
for $d_g = 1.1$ mm and $d_c=10$ mm, there is no change in slope at $H_c$.
The solid line has slope 2. 
}
\label{Fig4}
\end{figure}


\begin{references}
\bibitem{Plots}{
Unless otherwise noted, data are shown for $d_g = 0.9$ mm, $k = 25$ N/cm and
$v = 0.2$ mm/s.
}
\bibitem{Precursors}{
We see no evidence of precursors to the slips within our time resolution of
$\sim 0.2$ ms.
}
\bibitem{Couette}{
Similar scaling has been reported for sheared granular 
materials  in an annular Couette geometry \cite{BMiller}.
}
\bibitem{Beads}{
We have examined larger grains with average diameters 
of $1.4, 1.6, 3.0$ and $5.0$ mm .
}
\end{references}
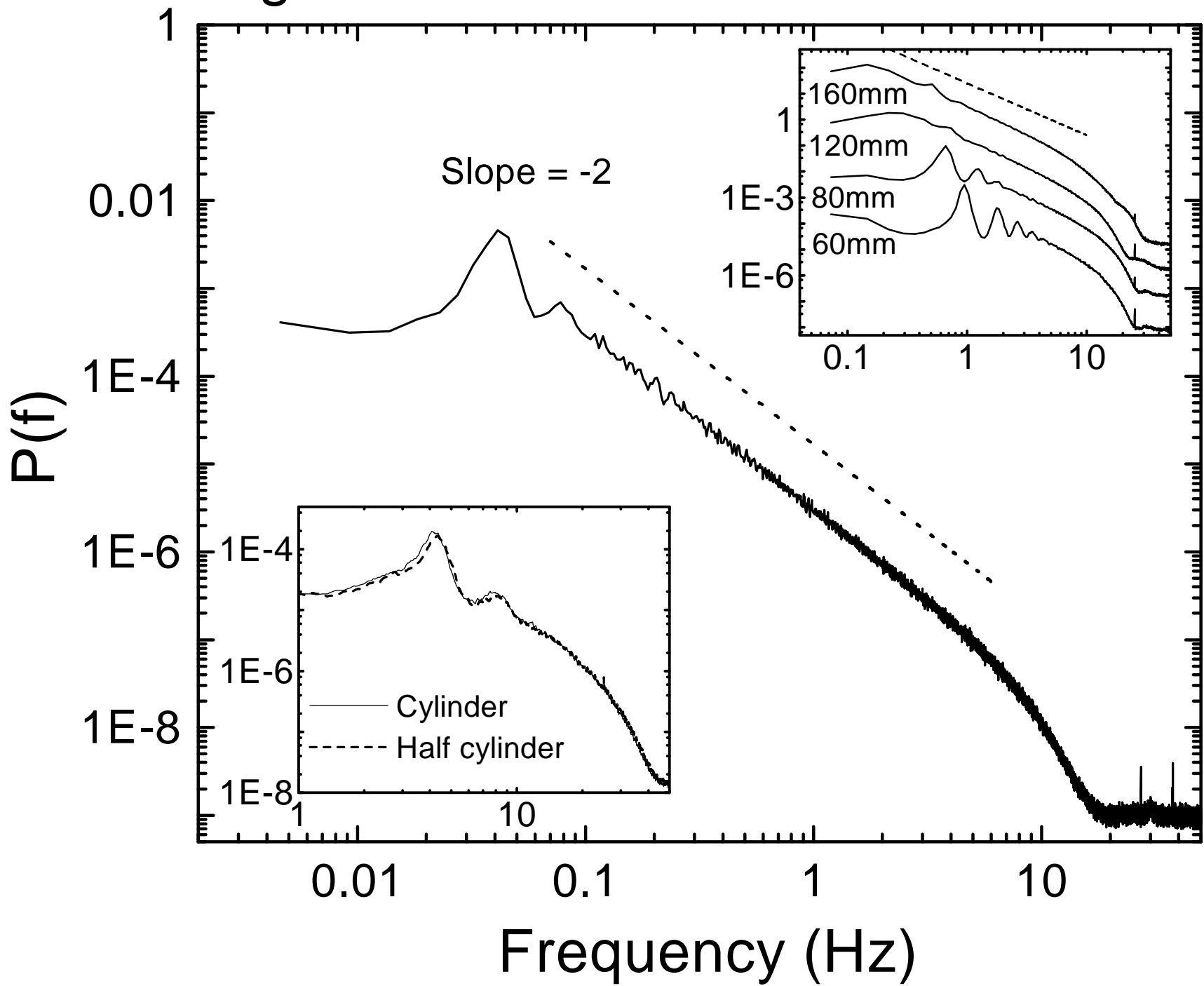
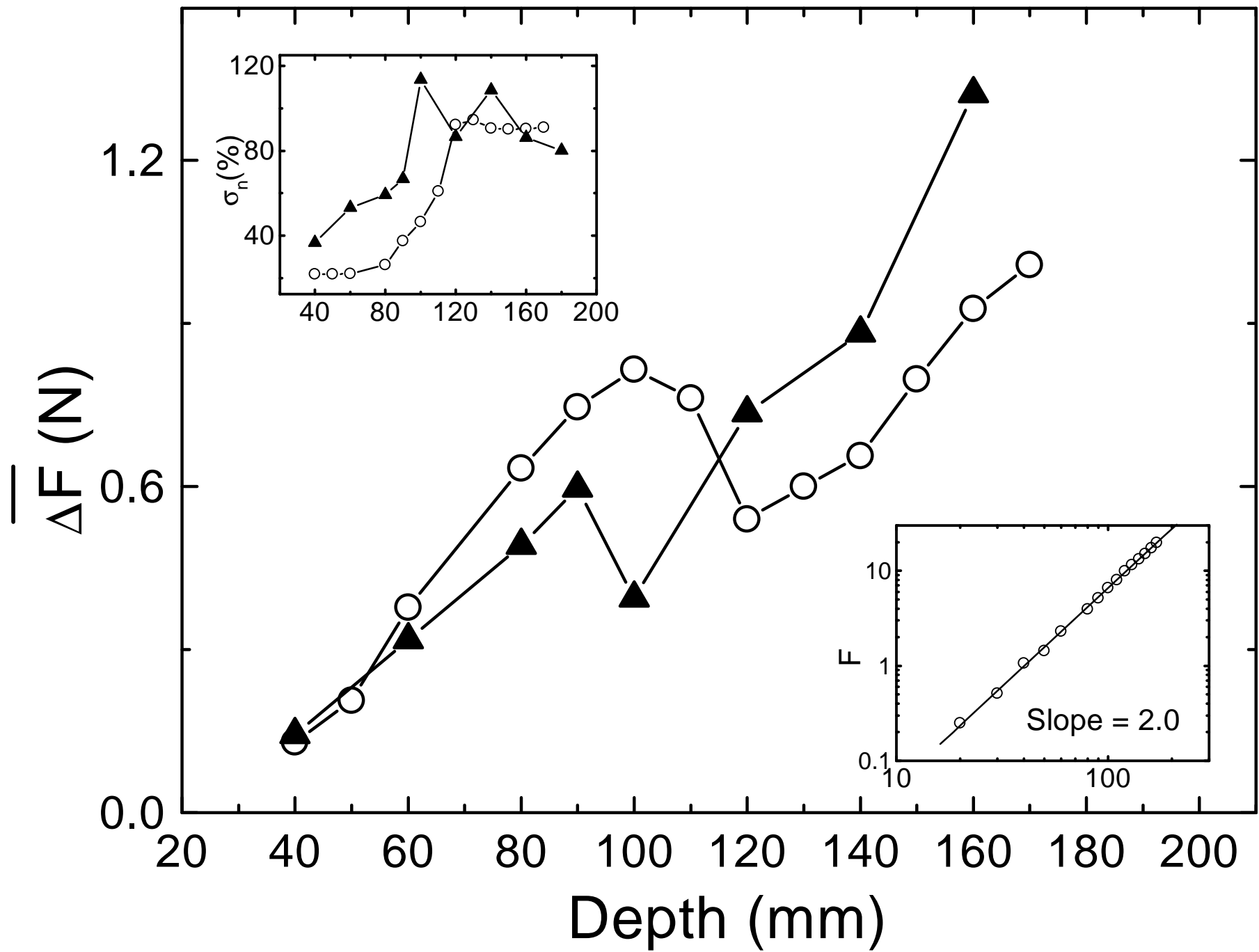
\end{document}